\documentclass[aps,prb,twocolumn,superscriptaddress,showpacs,floatfix]{revtex4}
\usepackage{graphicx}
\usepackage{amsmath}
\usepackage{amssymb}

\newcommand{\dmo}[4]{{#1}$_{#2}${#3}$_{#4}$MnO$_3$}

\graphicspath{{.}}

\begin{document}

\title{Interplay of charge, spin,  orbital and lattice 
correlations in\\ colossal magnetoresistance manganites} 

\author{Alexander Wei{\ss}e}
\affiliation{Physikalisches Institut, Universit\"at Bayreuth, 95440
  Bayreuth, Germany} 

\author{Holger Fehske}
\affiliation{Institut f\"ur Physik, Ernst-Moritz-Arndt Universit\"at
  Greifswald, 17487 Greifswald, Germany} 
\date{\today}

\begin{abstract}
  We derive a realistic microscopic model for doped colossal
  magnetoresistance manganites, which includes the dynamics of charge,
  spin, orbital and lattice degrees of freedom on a quantum mechanical
  level.  The model respects the $SU(2)$ spin symmetry and the full
  multiplet structure of the manganese ions within the cubic lattice.
  Concentrating on the hole doped domain ($0\le x\le 0.5$) we study
  the influence of the electron-lattice interaction on spin and
  orbital correlations by means of exact diagonalisation techniques.
  We find that the lattice can cause a considerable suppression of the
  coupling between spin and orbital degrees of freedom and show how
  changes in the magnetic correlations are reflected in dynamic phonon
  correlations.  In addition, our calculation gives detailed insights
  into orbital correlations and demonstrates the possibility of
  complex orbital states.
\end{abstract}

\maketitle

\section{Introduction}
The observation of the colossal magnetoresistance effect
(CMR)~\cite{KSKMH89,HWHSS93,JTMFRC94} in doped manganese oxides with
perovskite structure, \dmo{R}{1-x}{A}{x} (R = rare-earth, A =
alkaline-earth metal), moved these materials into the focus of intense
research activity~\cite{CVM99,TT99,DHM01}. It turned out soon that the
complex electronic and magnetic properties of the manganites depend on
a close interplay of almost all degrees of freedom known in solid
state physics, namely itinerant charges, localised spins, orbitals,
and lattice vibrations.
On the one hand, the strong Coulomb interaction $U$ and the Hund's rule
coupling $J_h$ introduce a spin background and affect the charge mobility
via double-exchange~\cite{Ze51b,AH55,KO72a}. On the other hand, the
cubic environment of the Mn sites within the perovskite lattice
results in a crystal field splitting of Mn-$d$-orbitals into $e_g$ and
$t_{2g}$ and gives rise to an orbital degeneracy in the ground-state
of Mn$^{3+}$ ions. This orbital degeneracy, in turn, connects the
electronic system to the lattice, making it sensible to Jahn-Teller 
distortion and polaronic effects~\cite{Mi98}.

There are numerous attempts to describe the electronic (i.e., charge,
spin, and orbital) interactions of the manganites on a microscopic
level (see, e.g., Refs.~\onlinecite{IIM96,SNS97,MIN98}). However, most
of these models do not reflect the full multiplet structure of the Mn
ions caused by the cubic site symmetry or violate the spin rotational
invariance.  To our knowledge, Feiner and Ol\'es~\cite{FO99} were the
first who derived a consistent spin-orbital model for the undoped
compounds. In the following we extend their derivation to the case of finite
doping and complete the resulting electronic model with dynamic
electron-lattice interactions of Jahn-Teller and breathing type.
Up to now models of similar complexity have only been discussed on the
basis of rather extensive approximations, where mean-field approaches
are probably the most popular. In particular, the lattice was always
treated within the adiabatic approximation which is questionable in view
of the comparable energy scales of spins and phonons. For a better
understanding of this complex many body system we prefer to consider
all interactions on an equal footing. Using exact diagonalisation
techniques we study the ground-state properties of the derived model
on a four site cluster.

For the hole doped region ($0\le x\le 0.5$) our calculations show how
the lattice can effectively control the spin and orbital correlations and
the charge mobility. In the undoped compounds the electron-phonon
interaction suppresses the coupling of spin and orbital degrees of
freedom and is most effective in determining orbital order.  At doping
$x=0.25$, where ferromagnetism is stabilised by the double-exchange
mechanism, the coupling to the lattice can cause self-trapping of the
charge carriers, which immediately switches the spin order. In turn,
this change is reflected in dynamic lattice correlations, namely the
bond length fluctuation. For the half-doped manganites we find that
the electron-phonon coupling enhances the susceptibility for charge
ordering which is present already due to the strong Coulomb
interaction. In addition, moderate variations in the strength of the
electron-lattice coupling trigger pronounced changes in spin and
orbital correlations. This relates to the complicated patterns of spin,
charge and orbital ordering observed experimentally for different
manganites at $x=0.5$.

\section{Microscopic model}

\subsection{Electronic interactions}
\begin{figure}[b]
  \begin{center}
    \includegraphics[width=0.7\linewidth]{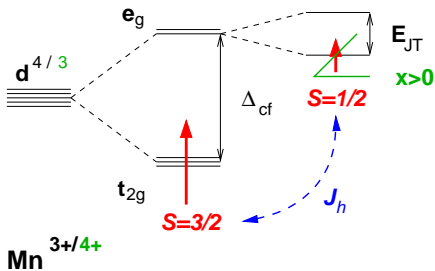}
  \end{center}
  \caption{Local electronic structure of Mn-$d$-electrons in a cubic
    environment. Here $\Delta_{\text{cf}}$, $J_h$, and $E_{\text{JT}}$
    denote the crystal field splitting, the Hund's rule coupling, and
    the Jahn-Teller splitting, respectively.} 
  \label{figlevels}
\end{figure}

The electronic and magnetic properties of the mixed-valence manganites
are governed by the manganese $d$-electrons, which are divided into
$t_{2g}$ and $e_g$ according to the cubic symmetry within the crystal
field. Due to the large Coulomb and Hund's rule interactions each of
the three $t_{2g}$ and the two $e_g$ levels carries at most one electron
and the spins of several $d$-electrons are aligned in
parallel at every site (see Figure~\ref{figlevels}).  
Starting from the undoped compounds with only Mn$^{3+}$
ions doping will remove $e_g$-electrons. Hence, the local electronic
Hilbert space can be restricted to the large Hund's rule ionic
ground-states of manganese ions in a cubic crystal field (see
Ref.~\onlinecite{Gr71}). For Mn$^{3+}$ ($d^4$) this corresponds 
to the spin-$2$ orbital doublet $^{5}E$ [$t_2^3(^4A_2)e$],
\begin{equation}\label{eq5ebase}
  \begin{aligned}
    |\theta,2,m\rangle & = +
    \,\sqrt{\tfrac{(2-m)!}{4!(2+m)!}}\,(S^{+})^{(2+m)}\,
    c^{\dagger}_{\varepsilon\downarrow} c^{\dagger}_{\xi\downarrow}
    c^{\dagger}_{\eta\downarrow} c^{\dagger}_{\zeta\downarrow}|0\rangle\\
    |\varepsilon,2,m\rangle & = -
    \,\sqrt{\tfrac{(2-m)!}{4!(2+m)!}}\,(S^{+})^{(2+m)}\,
    c^{\dagger}_{\theta\downarrow} c^{\dagger}_{\xi\downarrow}
    c^{\dagger}_{\eta\downarrow}
    c^{\dagger}_{\zeta\downarrow}|0\rangle\,,
  \end{aligned}
\end{equation}
and for Mn$^{4+}$ ($d^3$) to the spin-$\frac{3}{2}$ orbital singlet
$^{4}A_2$ [$t_2^3$],
\begin{equation}\label{eqa2base}
  |a_2,\tfrac{3}{2},m\rangle =
  \sqrt{\tfrac{(\frac{3}{2}-m)!}{3!(\frac{3}{2}+m)!}}\,
  (S^{+})^{(\frac{3}{2}+m)}\,
  c^{\dagger}_{\xi\downarrow} c^{\dagger}_{\eta\downarrow}
  c^{\dagger}_{\zeta\downarrow}|0\rangle\,.
\end{equation}
Here the operators $c^{\dagger}_{\alpha\sigma}$ create spin $\sigma$
electrons in the $e_g$ ($\alpha=\theta,\varepsilon$) or $t_{2g}$
($\alpha=\xi,\eta,\zeta$) orbitals and $S^+ = \sum_\alpha
c^{\dagger}_{\alpha\uparrow}c^{}_{\alpha\downarrow}$ raises the on-site
spin.
%
\begin{figure}
  \begin{center}
    \includegraphics[width=\linewidth]{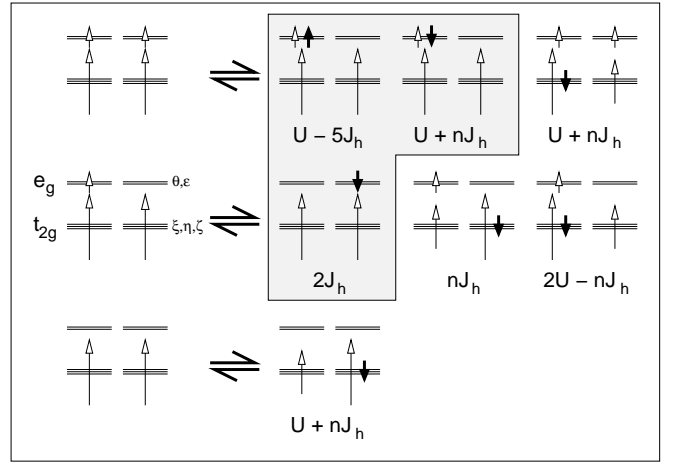}
  \end{center}
  \caption{Second order virtual excitations contributing to the model
    Hamiltonian $H_{\text{el}}$, Eq.~\eqref{hamel}. The shaded region
    corresponds to $t^2$ terms, the other terms are proportional to
    $t_{\pi}^2$. $n$ denotes different nonnegative prefactors of $J_h$
    (see Eq.~\eqref{eqelham}).}
  \label{figvirtual}
\end{figure}
The overlap of neighbouring manganese $d$ and oxygen
$p$ orbitals allows for a tunneling of the manganese electrons between
adjacent sites. Due to the specific symmetry of the involved
orbitals this hopping acquires an anisotropy~\cite{An59,KK72},
\begin{multline}
  H_{\text{t}} = - \sum_{i,\delta,\sigma}
  R_\delta\left[t\,c^{\dagger}_{i,\theta\sigma}
    c^{}_{i+\delta,\theta\sigma}\right.\\ + \left.  t_{\pi}
    (c^{\dagger}_{i,\xi\sigma} c^{}_{i+\delta,\xi\sigma} +
    c^{\dagger}_{i,\eta\sigma} c^{}_{i+\delta,\eta\sigma})\right] +
  \text{H.c.}\,,
\end{multline}
which can be expressed in terms of the spatial rotations
\begin{equation}
  \begin{gathered}
    R_x = (C_3^d)^{1}\,,\ R_y = (C_3^d)^{2} = (C_3^d)^{-1}\,,\
    R_z = (C_3^d)^{3} = 1\\
    C_3^d : c^{}_{\theta/\varepsilon}\rightarrow -\tfrac{1}{2}
    c^{}_{\theta/\varepsilon} \pm\tfrac{\sqrt{3}}{2}
    c^{}_{\varepsilon/\theta}\,;\ c^{}_{\xi/\eta/\zeta} \rightarrow
    c^{}_{\eta/\zeta/\xi}\,.
  \end{gathered}
\end{equation}
The transfer amplitudes $t$ and $t_{\pi}$ of $e_g$ and $t_{2g}$
electrons are small compared to the local energies $U$ and $J_h$.
Therefore the mobility of the $e_g$ electrons depends on the correlations
of the spin background formed by the $t_{2g}$ electrons. Their kinetic
energy is maximal, if neighbouring spins are aligned
ferromagnetically, which is the essence of the well known
double-exchange interaction~\cite{Ze51b,AH55}. Second order processes
in $t$ and $t_{\pi}$ are responsible for numerous superexchange
interactions between the localised spins, which inherit the orbital
anisotropy of the hopping.

The electronic part of our model is derived by second order degenerate
perturbation theory~\cite{FO99}, i.e., we calculated the matrix
elements of $H_{\text{t}}$ between the basis states~\eqref{eq5ebase}
or~\eqref{eqa2base} and all admissible excited states~\cite{Gr71}. In
Figure~\ref{figvirtual} the spin part of the fifteen different virtual
excitations contributing to the superexchange is summarised
graphically. We neglect terms which involve three different lattice
sites.

A compact expression for the resulting electronic Hamiltonian,
\begin{equation}\label{hamel}
  H_{\text{el}} =
  \sum_{i,\delta} R_{\delta}(H^{z}_{i,i+\delta}+\textrm{H.c.})\,,
\end{equation}
is obtained by rewriting the orbital and charge degrees of freedom of
the basis states~(\ref{eq5ebase}) and~(\ref{eqa2base}) in terms of new
Fermi operators $d^{\dagger}_{\alpha}$ and projectors $P^{\alpha}$,
\begin{align}
  |\theta\rangle & = d^{\dagger}_{\theta} |0\rangle, &
  |\varepsilon\rangle & = d^{\dagger}_{\varepsilon} |0\rangle, &
  |a_2\rangle & = d^{\dagger}_{\theta} d^{\dagger}_{\varepsilon}
  |0\rangle, \notag\\
  P^{\theta}_{} & = n^{}_{\theta}(1-n^{}_{\varepsilon}), &
  P^{\varepsilon}_{} & = n^{}_{\varepsilon}(1-n^{}_{\theta}), &
  P^{a_2}_{} & = n^{}_{\varepsilon} n^{}_{\theta}.
\end{align}
Note that these projectors are related to the common pseudo-spin
operators 
\begin{equation}
  \tau^{z} = \tfrac{1}{2}\sigma^{z}\,,\quad
  \tau^{x/y} = R_{x/y}(\tau^{z}) =
  \tfrac{1}{4}(-\sigma^{z}\mp\sqrt{3}\sigma^{x})\,,
\end{equation}
by the equation $R_{\delta}(P^{\theta/\varepsilon}) =
\tfrac{1}{2}\pm\tau^{\delta}$.  Since the amplitude of the on-site
spin is different for the basis states~(\ref{eq5ebase})
and~(\ref{eqa2base}), it is convenient to represent the spin degree of
freedom by Schwinger bosons $a^{}_{\uparrow}$ and $a^{}_{\downarrow}$
which allow for a uniform description. Then, the spin operator is
defined by $2\,{\bf S} =
a^{\dagger}_{\mu}\boldsymbol{\sigma}^{}_{\mu\nu} a^{}_{\nu}$, but the
amplitude is subject to the constraint $2\,|{\bf S}| =
a^{\dagger}_{\uparrow}a^{}_{\uparrow} +
a^{\dagger}_{\downarrow}a^{}_{\downarrow} = 4 -
n^{}_{\theta}n^{}_{\varepsilon}$.  Using this notation in
$z$-direction the interaction between nearest-neighbour sites is given
by,
\begin{widetext}
  \begin{multline}\label{eqelham}
    H^{z}_{i,j} =
    -\frac{t}{5}\left(a^{}_{i,\uparrow}a^{\dagger}_{j,\uparrow}
      +a^{}_{i,\downarrow}a^{\dagger}_{j,\downarrow}\right)
    \left(d^{\dagger}_{i,\theta} n^{}_{i,\varepsilon} d^{}_{j,\theta}
      n^{}_{j,\varepsilon}\right) + t^2\,\frac{{\bf S}_{i} {\bf
        S}_{j}-3}{32 J_h} \,P^{\varepsilon}_{i}P^{a_2}_{j} -
    t^2\,\frac{{\bf S}_{i} {\bf S}_{j}+6}{10(U-5J_h)}
    \,P^{\varepsilon}_{i}P^{\theta}_{j}\\
    + t^2\,\frac{{\bf S}_{i} {\bf S}_{j}-4}{8}
    \left[\frac{(4U+J_h)\,P^{\varepsilon}_{i}P^{\theta}_{j}} {5 U
        (U+\frac{2}{3} J_h)} +
      \frac{(U+2J_h)\,P^{\varepsilon}_{i}P^{\varepsilon}_{j}}
      {(U+\frac{10}{3} J_h)(U+\frac{2}{3} J_h)}\right] +
    t_{\pi}^2\,\frac{\frac{4}{9}{\bf S}_{i}{\bf
        S}_{j}-1}{U+\frac{4}{3}J_h}
    \,P^{a_2}_{i}P^{a_2}_{j}\\
    + t_{\pi}^2\,\frac{{\bf S}_{i} {\bf S}_{j}-3}{3}
    \left[\frac{(U-2J_h)(R_{x}(P^{\varepsilon}_{i}P^{a_2}_{j})+
        R_{y}(P^{\varepsilon}_{i}P^{a_2}_{j}))} {\frac{19}{3}J_h
        (2U-\frac{7}{3}J_h)} +
      \frac{(U+\frac{5}{3}J_h)(R_{x}(P^{\theta}_{i}P^{a_2}_{j})+
        R_{y}(P^{\theta}_{i}P^{a_2}_{j}))}{\frac{13}{3}J_h(2U-J_h)}
    \right]\\
    + t_{\pi}^2\,\frac{{\bf S}_{i} {\bf S}_{j}-4}{8}
    \left[\frac{R_{x}(P^{\varepsilon}_{i}P^{\varepsilon}_{j})+
        R_{y}(P^{\varepsilon}_{i}P^{\varepsilon}_{j})}{U+8J_h/3} +
      \frac{R_{x}(P^{\theta}_{i}P^{\theta}_{j})+
        R_{y}(P^{\theta}_{i}P^{\theta}_{j})}{U+2J_h} +
      \frac{(2U+\frac{14}{3}J_h)(R_{x}(P^{\varepsilon}_{i}P^{\theta}_{j})+
        R_{y}(P^{\varepsilon}_{i}P^{\theta}_{j}))}{(U+4J_h)(U+\frac{2}{3}J_h)}
    \right]\,.
  \end{multline}
\end{widetext}
The first order in $t$ corresponds to the well known double exchange
interaction~\cite{Ze51b,AH55,KO72a}, which the authors discussed in
detail recently~\cite{WLF01a}. The second order in $t$ and $t_{\pi}$
appears to be more involved, since the number of accessible virtual
excitations is rather large (see Figure~\ref{figvirtual}).  However,
in all cases it is basically the product of a Heisenberg-type spin
interaction and two orbital projectors.  The energies of the virtual
excitations depend, in general, on all three Racah
parameters~\cite{Gr71} $A$, $B$, and $C$. Applying the convention of
Ref.~\onlinecite{FO99} we define $U$ and $J_h$ in terms of
$d^4d^4\rightleftharpoons d^5(^4A_1)d^3$ and $d^4d^4\rightleftharpoons
d^5(^6A_1)d^3$ excitations.
Together with the approximation~\cite{TS54b} $C\approx 4B$ this
yields $U = A + 22B$ and $J_h = 6 B$. Typical estimates for these
energies are $U\approx6\dots 9$~eV and $J_h\approx 0.6\dots
0.8$~eV.~\cite{ZS90,BMSNF92,MF95,Quea98}

\subsection{Electron-phonon interaction}
\begin{figure}
  \begin{center}
    \includegraphics[width=0.9\linewidth]{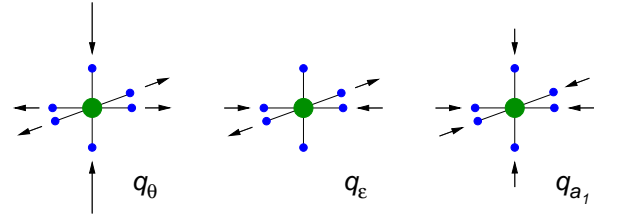}
  \end{center}
  \caption{Distortions of the oxygen octahedra for the Jahn-Teller
    ($q_\theta$, $q_\varepsilon$) and the breathing-type ($q_{a_1}$)
    phonon modes.}
  \label{figmodes}
\end{figure}

Concerning the electron-lattice interaction we concentrate on the
local environment of the manganese ions, namely the surrounding oxygen
octahedra (see Figure~\ref{figmodes}). At every site two optical
phonon modes of $E_g$ symmetry, $q_\theta$ and $q_\varepsilon$, couple
to the orbital degree of freedom of the $e_g$ electrons. In addition, a
breathing-mode $q_{a_1}$ is sensitive to the electronic density.  To
lowest order in the elongation $q_{\alpha} = b^{\dagger}_{\alpha} +
b^{}_{\alpha}$ ($\alpha\in\{\theta,\varepsilon,a_1\}$) this is
modelled by the $E\otimes e$ Jahn-Teller Hamiltonian~\cite{PW84}
\begin{multline}
  H_{\text{JT}}  = g\sum_i\left[
    (n^{}_{i,\varepsilon}-n^{}_{i,\theta})
    (b^{\dagger}_{i,\theta}+b^{}_{i,\theta})\right.\\
  +\ \left.(d^{\dagger}_{i,\theta} d^{}_{i,\varepsilon} +
    d^{\dagger}_{i,\varepsilon} d^{}_{i,\theta})
    (b^{\dagger}_{i,\varepsilon}+b^{}_{i,\varepsilon})\right]
\end{multline}
and a Holstein-type~\cite{Ho59a} interaction
\begin{equation}
  H_{\text{br}}  = \tilde g\sum_i (n^{}_{i,\theta} +
  n^{}_{i,\varepsilon}-2n^{}_{i,\theta}n^{}_{i,\varepsilon})
  (b^{\dagger}_{i,a_1}+b^{}_{i,a_1})\,.
\end{equation}
The bosonic operators $b^{}_{i,\alpha}$ describe the usual harmonic
lattice dynamics. To a good approximation we can assume these phonons to be
dispersion-less,
\begin{equation}
  H_{\text{ph}}  =  \omega\sum_i \left[b^{\dagger}_{i,\theta}b^{}_{i,\theta} +
    b^{\dagger}_{i,\varepsilon} b^{}_{i,\varepsilon}\right] +
  \tilde\omega\sum_i b^{\dagger}_{i,a_1}b^{}_{i,a_1}\,.
\end{equation}
In experiments~\cite{Ilea98,RB99} the vibrational modes involving
distortions of the oxygen octahedra are found to have frequencies of
the order of $50$ to $80$~meV. In the following we assume $g=\tilde g$
and $\omega =\tilde\omega$. This approximation reduces the number of
model parameters and is reasonable, since we are mainly interested in
qualitative features of the electron-lattice interaction.

\section{Numerical results}

\begin{figure}[h]
  \begin{center}
    \includegraphics[width=0.4\linewidth]{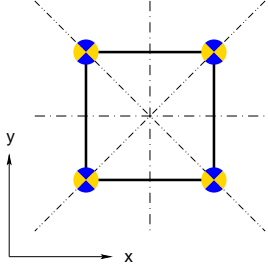}
  \end{center}
  \caption{Spatial symmetries of the considered cluster.}
  \label{figcluster}
\end{figure}

The Hilbert space of the complete microscopic model,
\begin{equation}\label{hamfull}
  H = H_{\text{el}} + H_{\text{JT}} + H_{\text{br}} + H_{\text{ph}}\,,
\end{equation}
is large and grows rapidly with the system size. However, using
a density matrix based optimisation procedure~\cite{WFWB00} for the
phonon subsystem we are able to retain the full quantum dynamics of
the lattice and the electronic subsystem in our numerical
calculation~\cite{CW85} of the ground-state properties. In addition we
take into account some of the discrete symmetries of the model, namely
the conservation of the $S^z$ component of the spin, the particle
number conservation, and the mirror symmetries orthogonal to the $x$ and $y$
axes (dot-dashed in Figure~\ref{figcluster}). Nevertheless, the typical
dimension of the eigenvalue problem is of the order of $10^6$ and its
repeated solution during the phonon optimisation requires the use of
large scale computers.

\subsection{Undoped manganites}
The undoped manganese oxides (\dmo{La}{}{}{}, \dmo{Pr}{}{}{}) usually exhibit
A-type anti-ferromagnetic order and strong Jahn-Teller distortion of
the ideal perovskite structure~\cite{WK55,Moea96}. The origin of the
observed magnetic order has been subject to discussions. While
different band structure calculations~\cite{Saea95,PS96a,SPV96,SHT96}
emphasise the importance of lattice distortions for the stability of
anti-ferromagnetism, Feiner and Ole\'s~\cite{FO99} favoured a purely
electronic mechanism.

\begin{figure}
  \begin{center}
    \includegraphics[width = \linewidth]{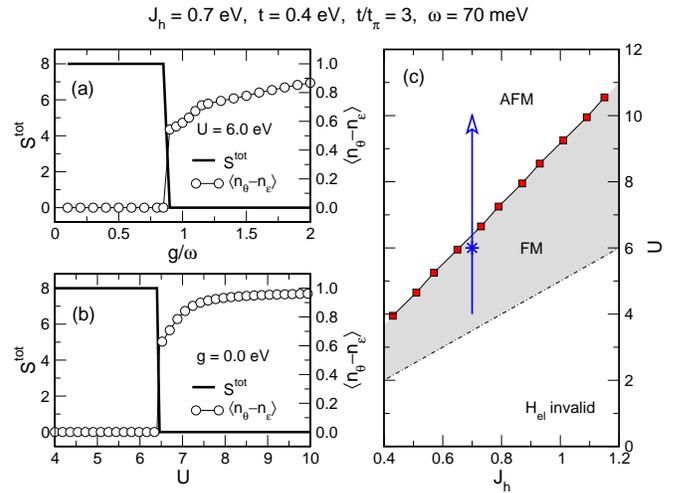}
  \end{center}
  \caption{Left: Total spin $S^{\text{tot}}$
    and orbital order ($\propto\langle n_\theta-n_\varepsilon\rangle$)
    of the ground-state of the cluster in dependence on (a) the 
    electron-phonon coupling $g$, and (b) the Coulomb interaction $U$. 
    Right: Phase diagram of the electronic model without
    electron-phonon interaction ($g=0$).}
  \label{figd0order}
\end{figure}

\begin{figure}
  \begin{center}
    \includegraphics[width = \linewidth]{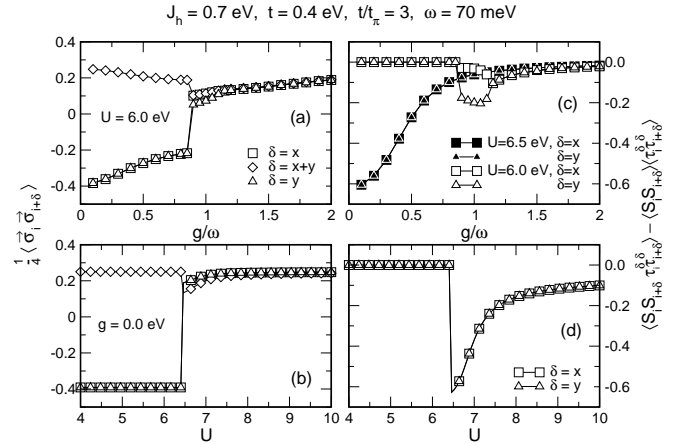}
  \end{center}
  \caption{Orbital-exchange [panels (a) and (b)] and spin-orbital
    [panels (c) and (d)] correlations between nearest and next-nearest
    neighbouring sites at variable electron-phonon coupling $g$ and
    Coulomb interaction $U$.}
  \label{figd0corr}
\end{figure}

Our calculation points out that both parameters, $U/J_h$ and $g$, can
drive a transition from ferromagnetic (FM) to anti-ferromagnetic (AFM)
order. Figure~\ref{figd0order}\,(c) shows the phase diagram of the
electronic model without electron-phonon interaction, i.e., $g=0$. We
assume $t=0.4$~eV and $t/t_{\pi}=3$ for the hopping
integrals~\cite{FO99} and characterise the magnetic phases according
to the total spin $S^{\text{tot}}$ of the ground-state of the four
site cluster. Parameters in the range $U\le 5J_h$ implicate that for
two neighbouring sites a $d^5d^3$ configuration becomes the
ground-state in favour of the $d^4(^5E)d^4(^5E)$ configuration. This
is incompatible with the situation in the manganites and consequently
the electronic Hamiltonian $H_{\text{el}}$ is not applicable for these values
of $U$ and $J_h$.  Starting from the FM phase with $5J_h\le U \lesssim
9.2J_h$ both, increasing $U$ or $g$, change the magnetic order of the
ground-state to AFM [Figure~\ref{figd0order}\,(a) and (b)]. The magnetic
transition is accompanied by a change in the corresponding orbital
order, which we identify by the local expectation value $\langle
n_\theta-n_\varepsilon\rangle$ and the orbital exchange correlation
$\langle\boldsymbol{\sigma}_{i}\boldsymbol{\sigma}_{i+\delta}\rangle$
between neighbouring sites. Here the Pauli matrices
$(\sigma^{\delta}_i)_{\mu\nu}$ operate on the orbital degree of
freedom, $\mu,\nu\in\{\theta,\varepsilon\}$, at the site $i$. The
panels~(a) and~(b) of the Figures~\ref{figd0order} and~\ref{figd0corr}
illustrate the transition from staggered to uniform orbital order.

In view of the distinct driving interactions ($U$ or $g$) both
transitions appear to be very similar. However, we observe a
significant difference, if we study the (de)coupling of spin and
orbital degrees of freedom. The latter has been a rather controversial
issue in the case of the Kugel-Khomskii model~\cite{KK73},
\begin{multline}\label{hamkuko}
  H = J\sum_{\langle ij\rangle_\alpha}\left[4({\bf S}_i {\bf
      S}_j)(\tau^\alpha_i+\frac{1}{2})(\tau^\alpha_j+\frac{1}{2})\right.\\ +
    \left.(\tau^\alpha_i-\frac{1}{2})(\tau^\alpha_j-\frac{1}{2}) - 1\right]\,,
\end{multline}
which contains the same kind of spin-orbital interactions ${\bf
  S}_{i}{\bf S}_{j}\,\tau^{\delta}_{i}\tau^{\delta}_{j}$ as our
Hamiltonian. Using equation of motion approaches for the approximate
solution of the model~\eqref{hamkuko} Khaliullin and
Oudovenko~\cite{KO97} decoupled spin and orbital degrees of freedom,
while Feiner, Ol\'es and Zaanen~\cite{FOZ98} emphasised the role of
mixed spin-and-orbital excitations. A correlation function, which makes
it possible to distinguish between these two decoupling schemes, is the
spin-orbital fluctuation $\langle{\bf S}_{i}{\bf S}_{i+\delta}\,
\tau^{\delta}_{i}\tau^{\delta}_{i+\delta}\rangle - \langle{\bf
  S}_{i}{\bf S}_{i+\delta}\rangle
\langle\tau^{\delta}_{i}\tau^{\delta}_{i+\delta}\rangle$. The data we
calculated for the complete manganite model~\eqref{hamfull} is given
in Figure~\ref{figd0corr}\,(c) and~(d). It indicates that this
correlation is smaller by a factor of $3$ to $5$, if phonons are
responsible for the FM to AFM transition (open symbols). On the other
hand, starting within the AFM phase increasing electron-phonon
interaction clearly suppresses the spin-orbital-fluctuation
[Figure~\ref{figd0corr}\,(c), bold symbols].  This behaviour is, of
course, crucial for effective theories that are based on such
decoupling schemes.

\begin{figure}
  \begin{center}
    \includegraphics[width = \linewidth]{wf02fig07.eps}
  \end{center}
  \caption{Behaviour of the lattice with increasing electron-phonon
    coupling $g$: (a) expectation value, and (b) fluctuation of the bond
    lengths $q_\delta$; (c) and (d) correlations of the Jahn-Teller
    modes $q_\theta$ and $q_\varepsilon$ between different sites.}
  \label{figd0latt}
\vspace{2mm}
  \begin{minipage}{0.3\linewidth}
    \begin{center}
      \includegraphics[width=\linewidth]{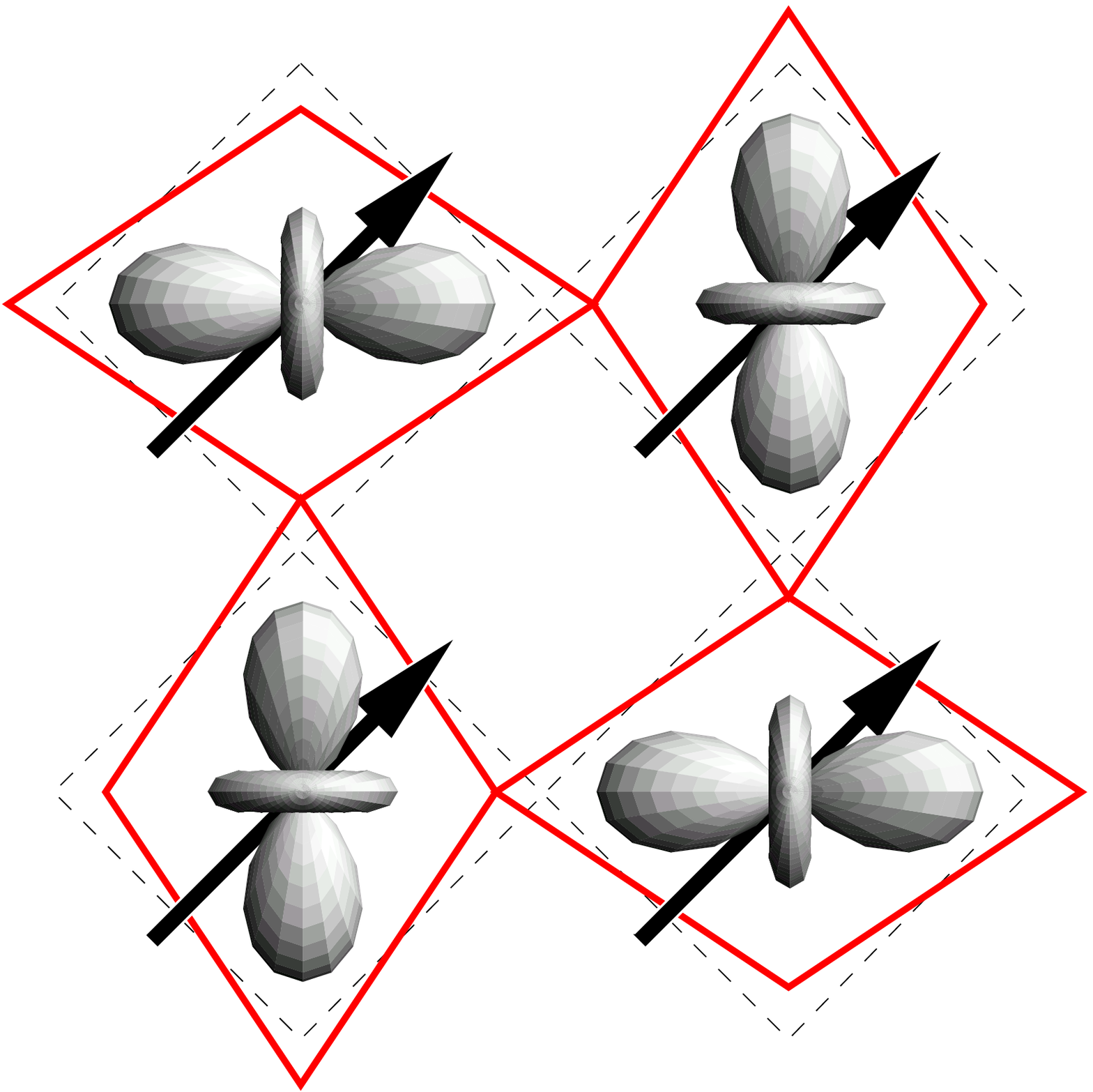}
    \end{center}
  \end{minipage}
  \begin{minipage}{0.3\linewidth}
    \begin{center}
      \includegraphics[width=\linewidth]{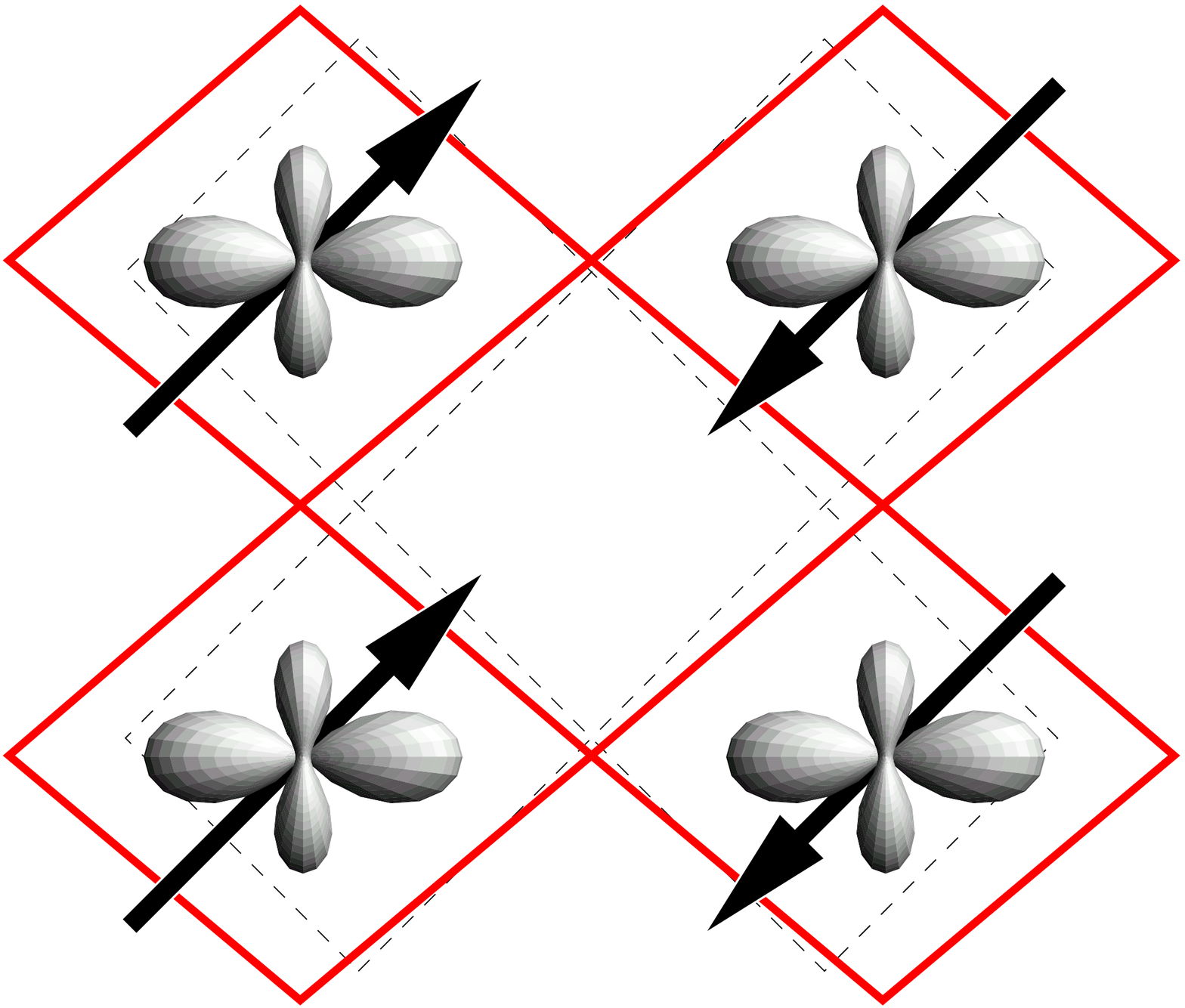}
    \end{center}
  \end{minipage}
  \begin{minipage}{0.3\linewidth}
    \begin{center}
      \includegraphics[width=\linewidth]{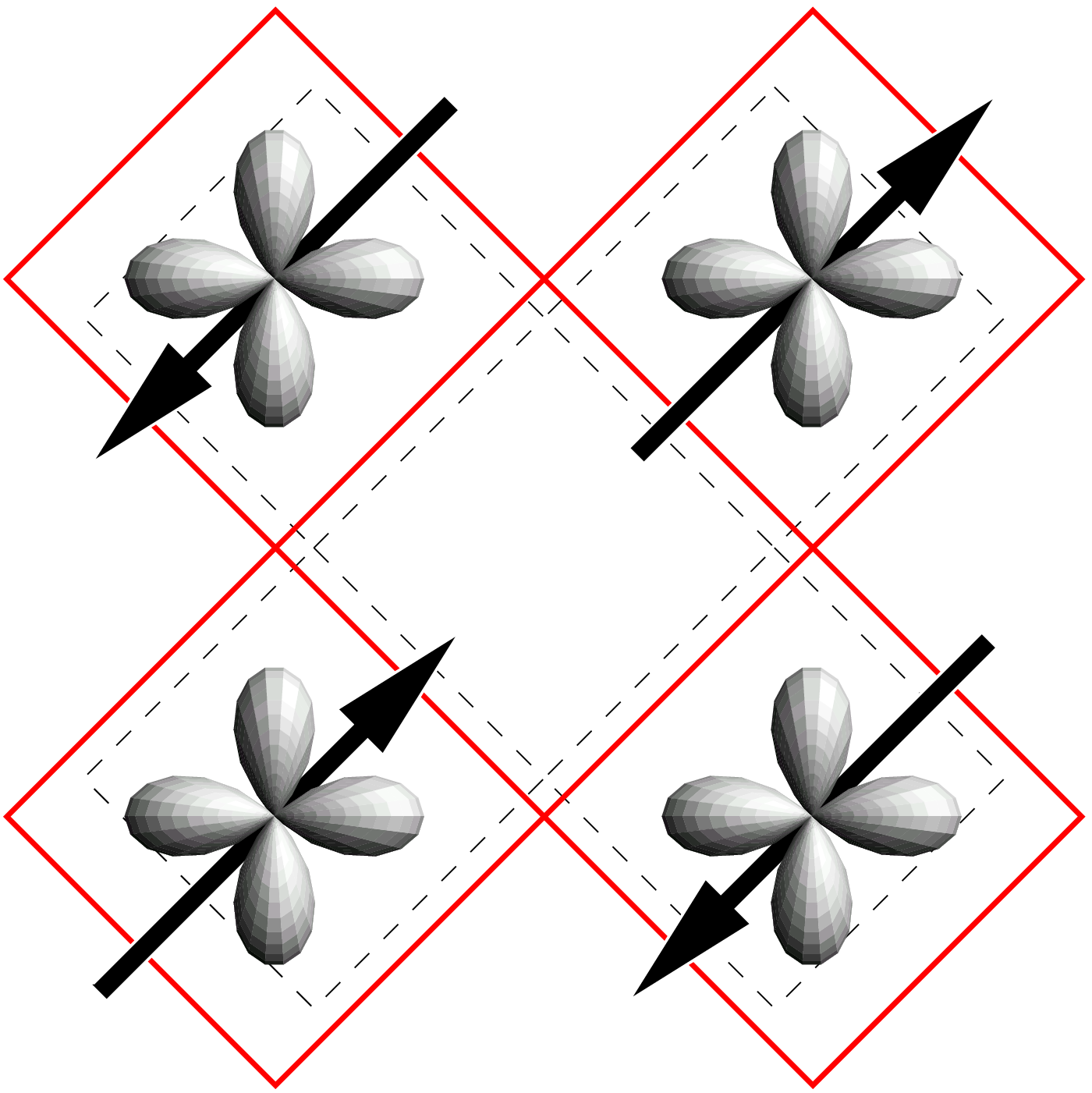}
    \end{center}
  \end{minipage}
  \begin{minipage}{0.9\linewidth}
    \begin{center}
      $\xrightarrow{\quad\text{increasing }g\quad}{}$
    \end{center}
  \end{minipage}
  \caption{Evolution of lattice, spin and orbital correlations with
    increasing electron-phonon coupling $g$ at doping $x=0$.}
  \label{figorbph0}
\end{figure}

To characterise the behaviour of the lattice we calculate the
correlations $\langle q_{i,\alpha} q_{j,\alpha}\rangle$
($\alpha\in\{\theta,\varepsilon\}$) between the elongations of the
Jahn-Teller modes $q_\theta$ and $q_\varepsilon$ at neighbouring sites
$i$, $j$. The bond lengths
\begin{equation}
  \begin{pmatrix} q_x \\ q_y \\ q_z \end{pmatrix} =
  \begin{pmatrix}
    \frac{1}{\sqrt{6}} & -\frac{1}{\sqrt{2}} & -\frac{1}{\sqrt{3}}\\
    \frac{1}{\sqrt{6}} & \frac{1}{\sqrt{2}} & -\frac{1}{\sqrt{3}} \\
    -\frac{2}{\sqrt{6}} & 0  & -\frac{1}{\sqrt{3}}
  \end{pmatrix}\cdot
  \begin{pmatrix} q_\theta \\ q_\varepsilon \\ q_{a_1} \end{pmatrix}
\end{equation}
define additional significant quantities. In particular we considered
the expectation values $\langle q_{i,\delta}\rangle$ and $\langle
q_{i,\delta}^2\rangle - \langle q_{i,\delta}\rangle^2$ with
$\delta\in\{x,y,z\}$. Figure~\ref{figd0latt} shows that within the FM
domain the lattice is undistorted but the $q_\varepsilon$ vibrations
are correlated along the $(\pi,\pi)$ direction in reciprocal space. At
intermediate values of $g/\omega\approx 1$ a finite $x$-$y$-distortion
develops, which is also reflected in spin and orbital correlations
[e.g., the spin-orbital fluctuation in Figure~\ref{figd0corr}\,(c)].
Such type of asymmetry seems surprising in view of the invariance of
the cluster with respect to diagonal reflections (dot-dot-dashed in
Figure~\ref{figcluster}). However, since this symmetry is not taken
into account in the calculation the system is trapped in one
particular linear combination of two degenerate ground-states and
different expectation values in $x$ and $y$ direction can evolve. At
larger electron-phonon coupling $g$ only a finite $q_\theta$
distortion remains and the $q_\varepsilon$ modes are uncorrelated.

In Figure~\ref{figorbph0} the change of orbital, spin and phonon
correlations with increasing $g$ is shown schematically. This, 
however, is a rather suggestive picture, which can not cover
all the details of the various correlations. To select the appropriate
graphical representation for the orbital arrangement we study the
$E\otimes E$-eigenstates of reduced orbital density matrices. On a
bond $\langle ij\rangle$ these states are either anti-symmetric,
\begin{equation}
  |a\rangle_{ij}\propto(|\theta\rangle_i\otimes|\varepsilon\rangle_j
  - |\varepsilon\rangle_i\otimes|\theta\rangle_j)\,,
\end{equation}
or symmetric, 
\begin{equation}
  |s(\varphi,\psi)\rangle_{ij}
  \propto(|\varphi\rangle_i\otimes|\psi\rangle_j 
  + |\psi\rangle_i\otimes|\varphi\rangle_j)\,,
\end{equation}
with respect to the permutation of two sites.  The angles
$\varphi,\psi\in\mathbb{C}$ parameterise two generalised orbital
states of the form $|\varphi\rangle = \cos(\varphi)|\theta\rangle +
\sin(\varphi)|\varepsilon\rangle$.  At small $g$ the staggered orbital
order implies that the anti-symmetric orbital state is the most
probable for each bond, i.e., it belongs to the largest eigenvalue of
the density matrix.  Since an anti-symmetric combination of two
arbitrary generalised states $|\varphi\rangle$ and $|\psi\rangle$ is
always proportional to $|a\rangle_{ij}$ the representation is
undefined. Therefore we took into account also the second eigenstate
of the density matrix, which is slightly less probable but necessarily
symmetric. The associated angles $\varphi$ and $\psi$ define the
orbital pattern shown in the left hand panel of
Figure~\ref{figorbph0}.  For larger $g$ the uniform orbital order
allows for a calculation of the depicted orbital pattern directly from
the most probable eigenstate of the density matrix. The spin
arrangements are chosen such as to reflect the expectation values of the
Heisenberg interaction, $\langle {\bf S}_i {\bf S}_j\rangle$, for
nearest neighbour bonds $\langle ij\rangle$.  Of course, the
calculated ordering patterns can not be directly compared to
experimental observations for the three dimensional compounds.
Nevertheless, the orbital arrangement within the FM spin background
closely resembles the data measured within the FM planes of the
A-type AFM structure~\cite{Muea98}.

\subsection{Doping $x=0.25$}

\begin{figure}
  \begin{center}
    \includegraphics[width = \linewidth]{wf02fig09.eps}
  \end{center}
  \caption{Doping $x=0.25$. (a) Total spin and kinetic
    energy vs. $g$; (b) Generalised orbital states $|\varphi\rangle$
    surrounding empty sites; (c) and (d) Expectation value $\langle
    q_\delta\rangle$ and fluctuation $\langle
    q_\delta^2\rangle-\langle q_\delta\rangle^2$ 
    of the bond length in $x$ and $y$ direction.}
  \label{figd1all}
\vspace{2mm}
  \begin{minipage}{0.3\linewidth}
    \begin{center}
      \includegraphics[width=\linewidth]{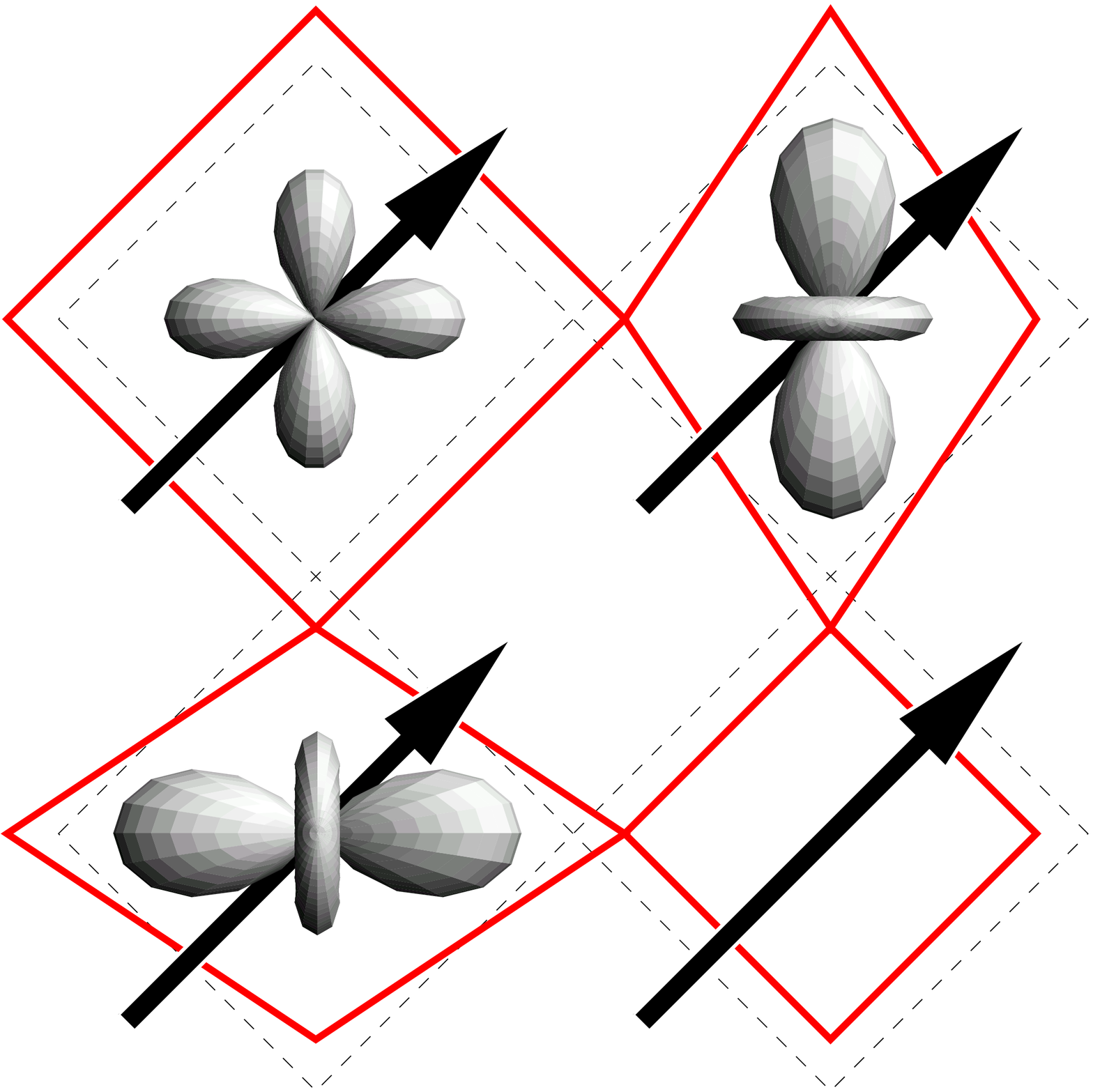}
    \end{center}
  \end{minipage}
  \begin{minipage}{0.3\linewidth}
    \begin{center}
      $\xrightarrow{\text{increasing }g}{}$
    \end{center}
  \end{minipage}
  \begin{minipage}{0.3\linewidth}
    \begin{center}
      \includegraphics[width=\linewidth]{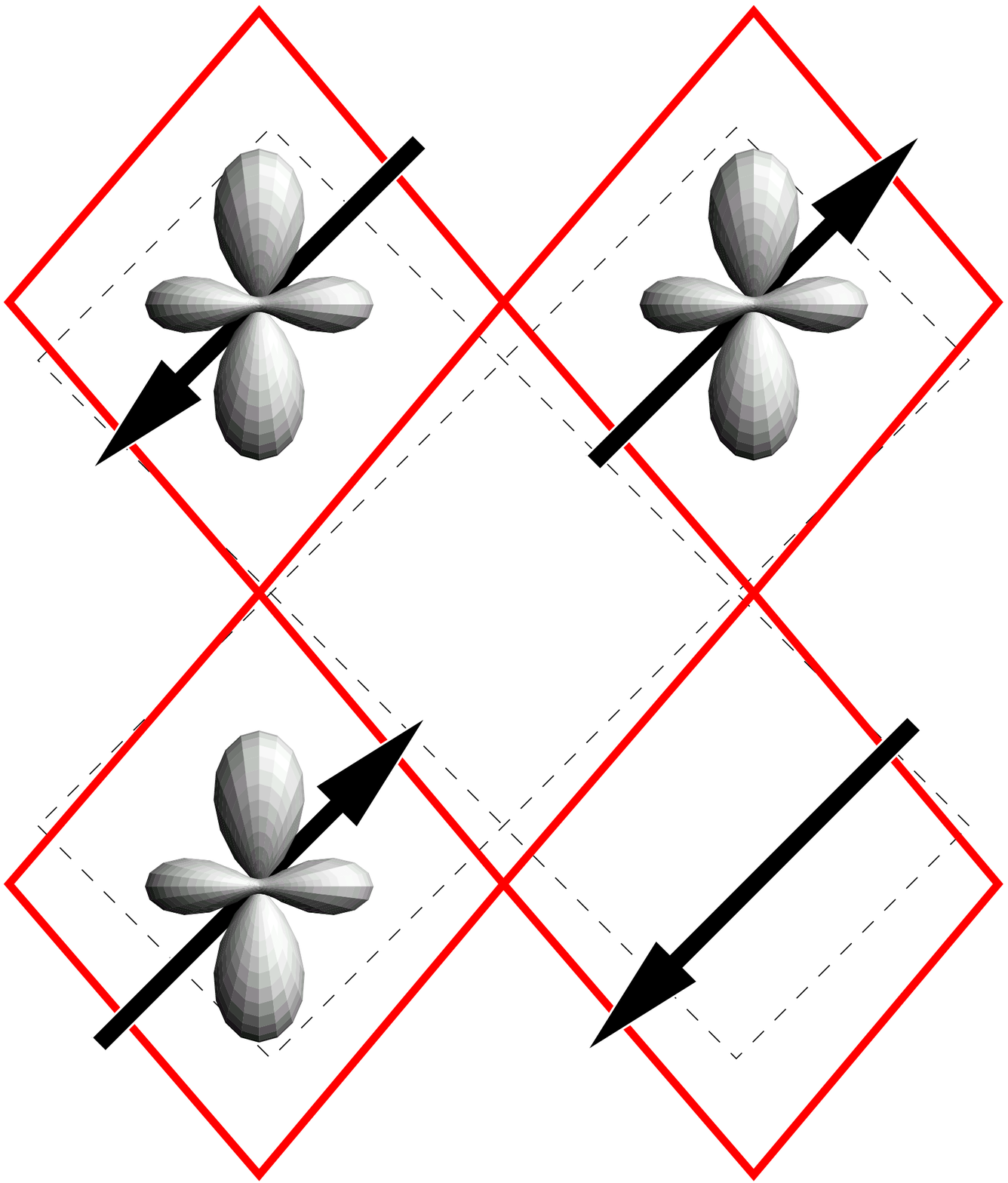}
    \end{center}
  \end{minipage}
  \caption{Evolution of lattice, spin and orbital correlations with
    increasing electron-phonon coupling $g$ at doping ${x=0.25}$.}
  \label{figorbph1}
\end{figure}
In view of the CMR effect the domain of low to intermediate doping
($0.15\le x\le 0.5$) is certainly the most interesting one. Here
ferromagnetism is stabilised by the double exchange interaction, which
depends on the mobility of the charge carriers. However, if too strong
electron-phonon coupling causes localisation of the holes, the spin
order breaks down.  For $x=0.25$ our calculation of the ground-state
properties clearly illustrates this coincidence.
Figure~\ref{figd1all}\,(a) shows the total spin $S^{\text{tot}}$ of
the cluster together with the expectation value of the kinetic energy
in the ground state,
\begin{equation}
  E_{\text{kin}} = \left\langle -\frac{t}{5}\sum_{i,\delta,\sigma} 
    R_{\delta}\!\left[
      a^{\phantom{\dagger}}_{i,\sigma}a^{\dagger}_{j,\sigma}
      \,d^{\dagger}_{i,\theta} n^{\phantom{\dagger}}_{i,\varepsilon}
      d^{\phantom{\dagger}}_{j,\theta}
      n^{\phantom{\dagger}}_{j,\varepsilon}\right] +
    \text{H.c.}\right\rangle\,.
\end{equation}
Obviously, the change from FM to AFM spin correlations with increasing
electron-phonon interaction $g$ is directly related to the charge
mobility. In addition the transition is accompanied by the appearance
of a lattice distortion in the $x$-$y$-plane [see
Figure~\ref{figd1all}\,(c)], which is interesting also in view of the
lattice dynamics. Whereas one of the elongations, $\langle
q_y\rangle$, grows linearly in $g$, the associated fluctuation
$\langle q_y^2\rangle-\langle q_y\rangle^2$ shows a kink near the FM
to AFM transition point. This behaviour closely reminds the data
collected by Booth~et~al.~\cite{BBKLCN98} using x-ray-absorption
fine-structure measurements (XAFS). For \dmo{La}{1-x}{Ca}{x} these
authors observed a similar rise in the Mn-O bond length variance
$\sigma^2$ near the critical temperature of the transition from the
ferromagnetic metallic to the paramagnetic insulating phase.

Orbital correlations are again extracted from the eigenstates of the
reduced density matrices for a single bond. We concentrate on the
environment of a hole, which is characterised by the $E\otimes A_2$
eigenstates, i.e., by states describing a Mn$^{3+}$ site neighbouring
a Mn$^{4+}$ site. Each of these states can be understood as the
product $|\varphi\rangle_i\otimes|a_2\rangle_j$ of a generalised
orbital state $|\varphi\rangle$ and the basis state $|a_2\rangle$ [see
Eq.~\eqref{eqa2base}]. In Figure~\ref{figd1all}\,(b) the corresponding
angle $\varphi$ of the most probable eigenstate of the density matrix
is given for bonds in $x$ and $y$ direction. The data for weak and
strong electron-phonon coupling $g$ is translated into the orbital
arrangement presented in Figure~\ref{figorbph1}. Increasing~$g$
reduces nearest-neighbour correlations and forces the orbital order to
depend locally on the dominant electron-phonon interaction. The
orbital polaron~\cite{KK99} pattern disappears, if the charge carrier
is trapped completely.

\subsection{Doping $x=0.5$}

\begin{figure}
  \begin{center}
    \includegraphics[width = \linewidth]{wf02fig11.eps}
  \end{center}
  \caption{Doping $x=0.5$. (a) Total spin and kinetic
    energy vs. $g$; (b) Complex orbital states on diagonal
    bonds; (c) Density-density correlations, (d) 
    Lattice distortion $\langle q_\delta\rangle$ in $x$ and $y$ direction.}
  \label{figd2all}
\vspace{2mm}
  \begin{minipage}{0.3\linewidth}
    \begin{center}
      \includegraphics[width=\linewidth]{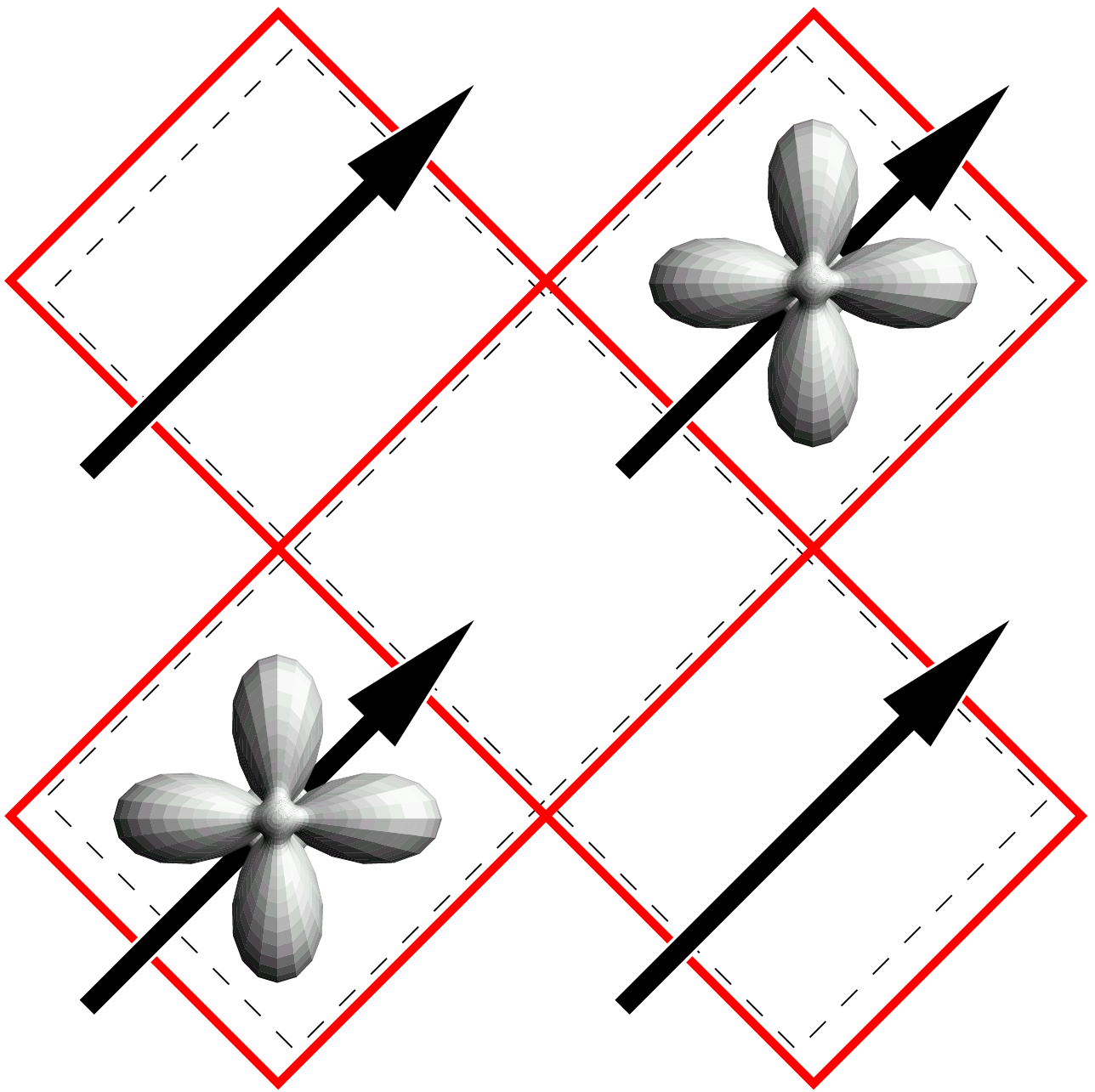}
    \end{center}
  \end{minipage}
  \begin{minipage}{0.3\linewidth}
    \begin{center}
      \includegraphics[width=\linewidth]{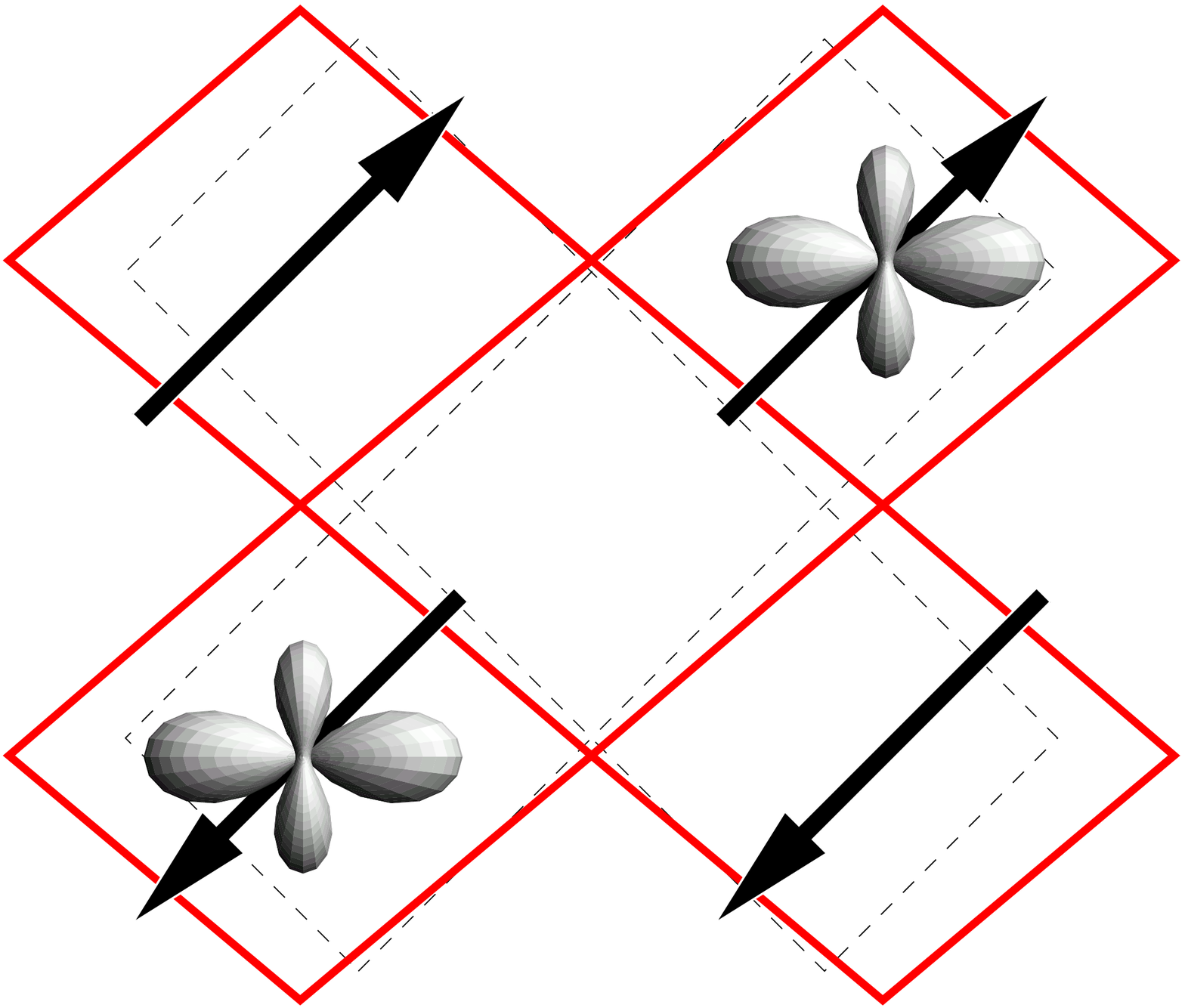}
    \end{center}
  \end{minipage}
  \begin{minipage}{0.3\linewidth}
    \begin{center}
      \includegraphics[width=\linewidth]{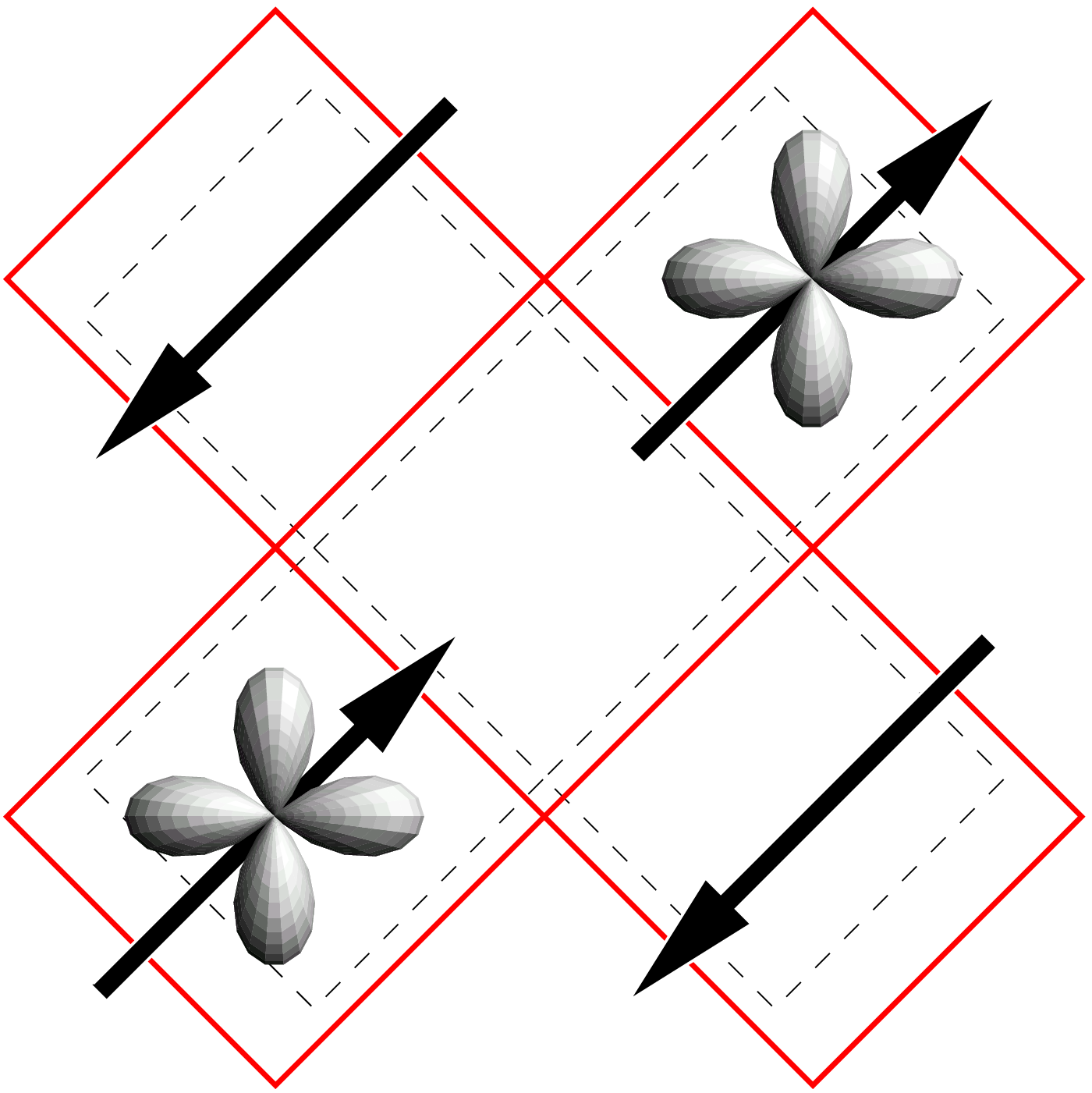}
    \end{center}
  \end{minipage}
  \begin{minipage}{0.9\linewidth}
    \begin{center}
      $\xrightarrow{\quad\text{increasing }g\quad}{}$
    \end{center}
  \end{minipage}
  \caption{Evolution of lattice, spin and orbital correlations with
    increasing electron-phonon coupling $g$ at doping $x=0.5$.}
  \label{figorbph2}
\end{figure}
At doping $x=0.5$ the picture is more involved. Again we consider
model parameters, which yield ferromagnetic spin correlations for
vanishing electron-phonon coupling, $g=0$. The strong Coulomb
interaction $U$ then leads to the formation of a charge density wave with
ordering vector $(\pi,\pi)$ in reciprocal space. Increasing electron-phonon
coupling $g$ amplifies this trend, and at large $g$ the trapping of the
carriers gives rise to charge order. In Figure~\ref{figd2all}\,(c)
this is illustrated with the density-density correlation between
different lattice sites $i$ and $j$. From the model
Hamiltonian~\eqref{eqelham} it is obvious that if charges tend to
maximise their mutual distance the particular anti-ferromagnetic
component of the Heisenberg spin interaction, which is proportional to
$t^2/J_h$, gains importance. Consequently FM order is unstable at much
lower~$g$, if compared to the case $x=0.25$. The FM to AFM transition is
not connected to charge localisation and causes only a tiny jump of
the kinetic energy [see Fig.~\ref{figd2all}\,(a)].

The calculated orbital correlations are interesting as well. On a
diagonal bond the most probable $E\otimes E$-eigenstate of the reduced
orbital density matrix is symmetric for all considered values of $g$,
$|s(\varphi,\psi)\rangle$. Note however, that the associated angles
are {\it complex} conjugated $\varphi = \psi^{*}\in\mathbb{C}$ within both
the FM and parts of the AFM phase [see
Figure~\ref{figd2all}\,(b)].  Here exact diagonalisation supports
recent mean-field studies~\cite{Kh00,TS00,MN00,BK01}, which obtain
different patterns of complex orbital ordering, in particular within the
ferromagnetic conducting phase. At large $g$ charge localisation
promotes real orbital states.

In Figure~\ref{figorbph2} we summarise the behaviour of the lattice,
the spins and the orbitals. With growing $g$ the cluster expands
isotropically until a finite $x$-$y$-distortion develops at the FM to AFM
transition [see also Figure~\ref{figd2all}\,(d)]. Further increase of $g$
causes the carriers to localise and the distortion to
disappear. The spin background undergoes a change from FM order to
different types of AFM order. Of course, within the four site system
we are not able to detect more complicated spin arrangements like the
CE-type order observed experimentally~\cite{WK55}. However, the
sensitivity of the system to small changes in the model parameters is
clearly visible. Note that the depicted orbitals represent the
amplitude of the underlying complex states $|\varphi\rangle$.

\section{Conclusion}
In the present work we have derived a microscopic model for doped CMR
manganites which includes the dynamics of charge, spin, orbital, and
lattice degrees of freedom on a quantum mechanical level. Using exact
diagonalisation techniques we have studied how the electron-lattice
interaction affects short range spin and orbital correlations. An
observation, which is important for the understanding of the undoped
compounds, is the suppression of the spin-orbital coupling with
increasing electron-phonon interaction.  For the weakly doped
compounds we have demonstrated the direct relationship between the trapping
of charge carriers and the breakdown of ferromagnetism. In addition, we have
shown that changes in the spin correlations are reflected in dynamic
lattice correlations. At intermediate doping we find that the system
depends on a subtle balance of double-exchange, superexchange and
electron-lattice interaction. The latter enhances the tendency for
charge ordering which in turn affects spin and orbital order. Besides
the calculation proves that complex orbital states can be a suitable
approximation for the description of orbital correlations.  Although
the calculated data is not quantitatively comparable to real,
three dimensional materials, the exact diagonalisation of even a small
system provides detailed insight into correlations and driving
interactions behind the rich phase diagram of the manganites. In
addition, the exact results may support the development of approximate
theories.

We thank L.F.~Feiner, F.~G\"ohmann, D.~Ihle, G.~Khaliullin, and
J.~Loos for valuable discussions. The grant of computational resources
by HLR~Stuttgart and NIC~J\"ulich, as well as financial support by the
Deutsche Forschungsgemeinschaft and the Czech Academy of Sciences is
acknowledged.

\end{document}